\documentclass[english,prl,twocolumn,showpacs,amsmath,amssymb,floatfix,superscriptaddress]{revtex4}

\usepackage{epsfig}
\usepackage{graphicx}
\usepackage{pdfsync}
\usepackage{ulem}
\usepackage{color}
 
\usepackage{hyperref} %do NOT move this!
\begin{document}

\title{Strongly interacting bosons in a 1D optical lattice at incommensurate densities}

\author{A. Lazarides}
\affiliation{Max Planck Institute for the Physics of Complex Systems, Noethnitzer Str.38, D-01187 Dresden, Germany}

\affiliation{Institute for Theoretical Physics, Utrecht University, Leuvenlaan 4, 3584CE Utrecht, the Netherlands}

\author{O. Tieleman}

\author{C. Morais Smith}
\affiliation{Institute for Theoretical Physics, Utrecht University, Leuvenlaan 4, 3584CE Utrecht, the Netherlands}

\date{\small\it \today}
\newcommand{\xip}{\ensuremath{\xi_p}}
\newcommand{\ignore}[1]{}
\newcommand{\tred}[1]{{\color{red} #1}}
\newcommand{\tblue}[1]{{\color{blue} #1}}

\begin{abstract}
We investigate quantum phase transitions occurring in a system of strongly interacting ultracold bosons in a 1D optical lattice. After discussing the commensurate-incommensurate transition, we focus on the phases appearing at incommensurate filling.  We find  a rich phase diagram, with superfluid, supersolid and solid (kink-lattice) phases.  Supersolids generally appear in theoretical studies of systems with long-range interactions; our results break this paradigm and show that they may also emerge in models including only short-range (contact) interactions, provided that quantum fluctuations are properly taken into account.
\end{abstract}


\maketitle

%\section{Introduction}  

The rapid progress in trapping and cooling atoms has rendered the study of ``tailor-made'' low-dimensional (D) systems~\cite{cazreview} experimentally accessible. Both the dimensionality and the interactions can be controlled, allowing great flexibility in realizing almost arbitrary strongly-correlated physical systems. A superfluid-Mott insulator (SF-MI) quantum phase transition, driven by increasing the potential depth of the optical lattice (and hence the relative strength of interactions) beyond a critical value, has been observed for bosons loaded into an optical lattice in 3D \cite{Bloch}, 2D \cite{Spielman}, and 1D \cite{Esslinger}. In addition, the Tonks-Girardeau gas, where bosons avoid spatial overlap and acquire fermionic properties due to strong repulsive interactions, has been experimentally realized in 1D \cite{TG}.  

Recently, a new type of quantum phase transition was observed in 1D in the very strongly interacting regime: for an arbitrarily weak optical lattice potential {\it commensurate}  with the atomic density of the Bose gas, a quantum phase transition into an insulating, gapped state, was observed, with the atoms pinned at the lattice minima~\cite{Chris}.  Theoretical studies of 1D systems based on the sine-Gordon model indeed predict that above a critical interaction strength, the SF should become a MI even for a {\it vanishingly weak} optical lattice \cite{Buchler}. 

In this Letter, we show that another interesting regime can be reached if the density is {\it incommensurate} with the optical lattice. The system is then described by a driven sine-Gordon model. In this model, the appearance of superfluidity (off-diagonal correlations) may be driven in two different ways, either by tuning the interaction strength at constant lattice depth and commensurate  period, as already realized experimentally~\cite{Chris}, or by tuning the density or lattice parameter away from commensurability. 
\begin{figure}[t]
	\centering
		\includegraphics[width=8cm]{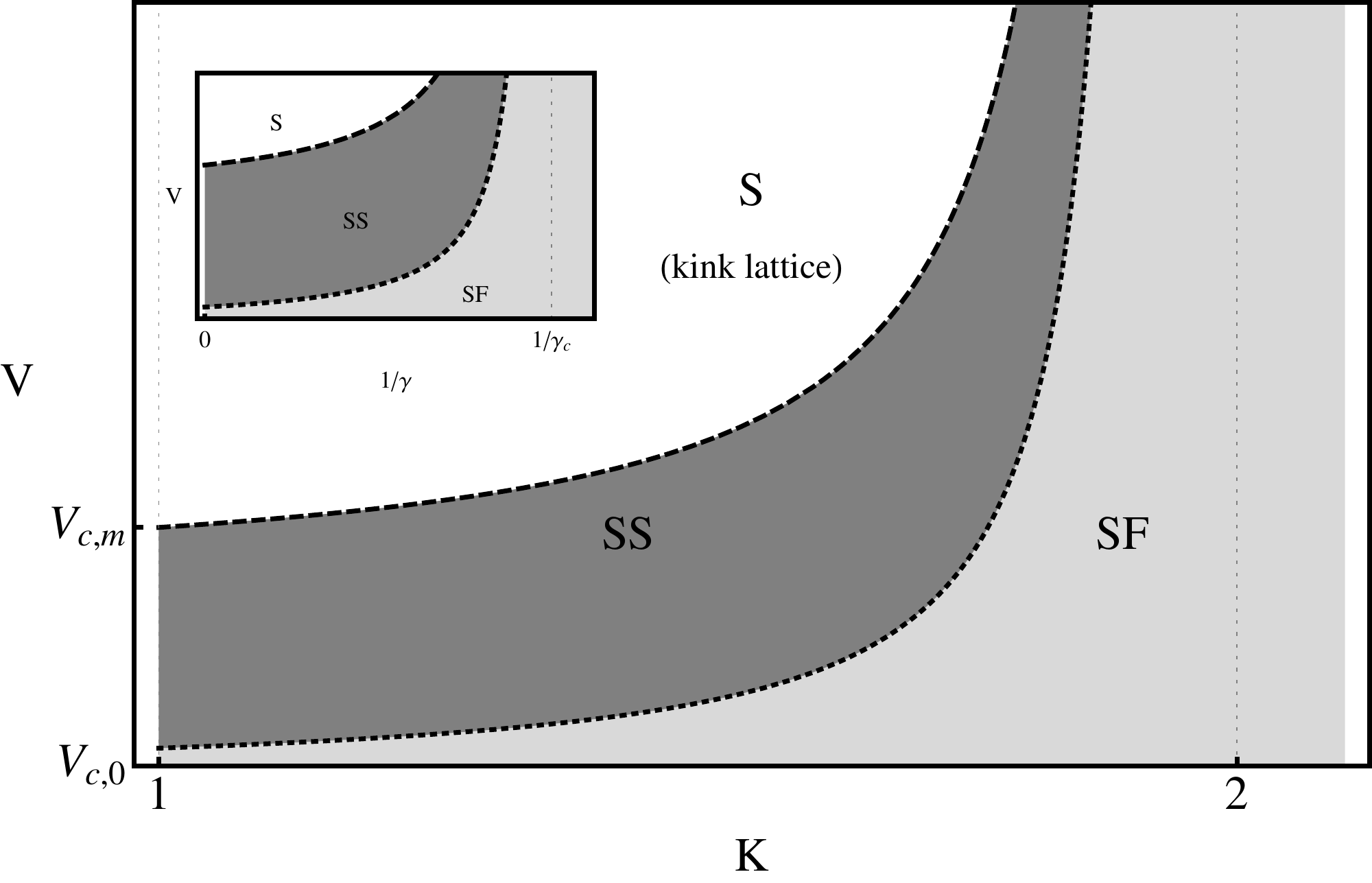}
	\caption{$T=0$ phase diagram for a filling factor slightly above unit filling. SF, SS, and S indicate superfluid, supersolid and solid phases, respectively. The inset shows the same phase diagram but in terms of $\gamma$, the dimensionless interaction strength, instead of the Tomonaga-Luttinger parameter $K$.}
	\label{figphasediagT0}
\end{figure}
We study the excitations of the system in the incommensurate phase and show that a supersolid (SS) phase may arise. In addition, for sufficiently large lattice strengths, a solid (S) phase is stabilized even at non-unit filling.

Our approach to studying this system is as follows: First, following previous work~\cite{Haldane,Buchler,Chris}, we formulate the underlying 1D interacting boson problem in terms of a quantum sine-Gordon field theory, with the deviation of the number density from commensurate values driving the appearance of kinks in the field. We carefully obtain the density threshold for the commensurate-incommensurate MI-SF transition, finding that the SF appears for {\it arbitrarily} small deviations of the density from the commensurate density, in agreement with Ref.~\cite{giamarchi}.

We next study the incommensurate regime, in which the excess particles appear as kinks of the sine-Gordon field, having an effective mass and effective interactions different from the bare particles. We extract these two parameters exactly from the underlying field theory and finally apply a functional renormalization group (RG) approach to the path integral formulation of the many-body statistical density matrix to obtain the ground state properties of the system. The RG transformation shows that quantum fluctuations renormalize the interactions between the kinks to a power law form; this maps the system to the Calogero-Sutherland model~\cite{beautifulmodels}, an exactly solvable model of 1D particles with long-range interactions. This finally allows us to propose a phase diagram for the incommensurate regime (see Fig.~\ref{figphasediagT0}). When the lattice potential is strong enough, the system solidifies. However, the S phase here is actually a lattice of kinks, and the number of particles per site is not fixed. At intermediate values of the lattice potential, we predict the emergence of a SS phase. SS phases usually occur in model Hamiltonians which include long-range interactions,  and have a characteristic wavelength which is an integer multiple of the lattice spacing~\cite{SS}. The most striking feature of the system studied in this Letter is that while the original Hamiltonian contains only {\it local} interactions, the SS phase appears due to the finite-range nature of the interaction between the excitations. In addition, the periodicity of the SS phase found here is {\it unrelated} to that of the lattice, a qualitatively different behavior from the situations usually found in the literature. 

%\section{Model} 
The microscopic description of a trapped gas of cold bosons in 1D with contact interactions and in the presence of a single-particle potential $V(x)$ is
%\begin{widetext}
%	\begin{equation}
\begin{eqnarray} \nonumber
	H&=&\int_{-\infty}^\infty dx\,
		\left[
			\psi ^\dagger (x)	
			\left(
				-\frac {\hbar^2}{2m} \nabla^2+V(x)
			\right)
			\psi (x) \right. \\  
			&+& \left.
			\frac {g}{2}
			\int_{-\infty}^\infty dx\,
			\psi ^\dagger (x)\psi ^\dagger (x)\psi(x)\psi(x)
		\right], 
	\label{eq:Hmicroscopic}
\end{eqnarray}
%	\end{equation}
%\end{widetext}
where $g$ is the strength of the $\delta$-function interaction, $\psi$ $(\psi^\dagger)$ are bosonic annihilation (creation) operators, and $m$ is the atomic mass. The parameter characterizing the strength of the interactions is the Lieb-Liniger parameter $\gamma=mg/\hbar^2n_0$, where $n_0$ is the average density.

Writing $\psi(x)=\sqrt{n(x)}\exp(-i\theta(x))$, with $n(x)$ the density and $\theta(x)$ the (real) phase, and using the Poisson summation formula, the density operator may then be expressed as~\cite{Haldane,Buchler} 
\begin{equation}
	n (x)=\left(n_0-\frac{1}{\pi}\partial_x\phi(x)\right)
			\sum_{p=-\infty}^\infty 
			e^{2ip[\pi n_0 x-\phi(x)]},
			\label{eqrhointermsofphi}
\end{equation}
where $\int dx\partial_x \phi(x)=0$. Eq.~\eqref{eqrhointermsofphi} yields an expression for the bosonic operators in Eq.~\eqref{eq:Hmicroscopic} in terms of the new field $\phi$. The appropriate bosonic commutation relations are satisfied if $\left[\partial_x\phi(x),\theta(x')\right]=-i\pi\hbar\delta(x-x')$; that is, $\theta$ and $\partial_x\phi/\pi$ are canonically conjugate variables. From Eq.~\eqref{eqrhointermsofphi}, it follows that {\it kinks in the $\phi$ field correspond to particle-like excitations}. This fact will be of great importance to us later on.

In the long-wavelength limit, and in the presence of an optical lattice creating a single-particle potential $V(x)=(V/2) \cos(4\pi x/\lambda_L)$, the system of Eq.~\eqref{eq:Hmicroscopic} may be described by an action of the form
\begin{eqnarray} \nonumber 
\mathcal{S}[\phi] &=&\int_0^\beta d\tau
	\int_{-\infty}^\infty dx\,
	\frac {1}{8\pi K}
	\left[
		\left(\partial_x\phi\right)^2
			+
		\left(\partial_\tau\phi\right)^2
	\right] \\ 
	&+&
	\frac {1}{2} u
	\int_0^\beta d\tau\int_{-\infty}^{\infty} 
	dx\,\cos\left[\phi(x)-Q x\right],
\label{eqaction}
\end{eqnarray}
where we have now set $\hbar=1$, scaled lengths such that the speed of sound is unity, and finally scaled $\phi\rightarrow\phi/2$. Here, $\beta=1/k_B T$, $u=n_0 V$, while $K$ is the usual Luttinger liquid parameter. For bosons interacting via contact potentials, $K$ may be expressed in terms of $\gamma$; for large $\gamma$, $K\approx (1+2/\gamma)^2$, while for smaller interaction strengths $\gamma$ it is given by $K\approx 
\pi/\sqrt{\gamma-\gamma^{3/2}/(2\pi)}$. Notice that $K\geq 1$, as it should for bosons with local interactions. We have also only kept the most relevant (least quickly oscillating in space) terms and written $Q=2\pi\left(n_0-2/\lambda_L \right)$ as the deviation of the average density from its commensurate value.

In the zero-temperature, $\beta\rightarrow\infty$ limit, Eq.~\eqref{eqaction} is formally equivalent to the model studied in Ref.~\cite{Lazarides2009}. It is also related to previous work on quantum Hall bilayer systems~\cite{Read,Tieleman}, with the important difference that the boundary conditions in the present case are $\int dx\, \partial_x\phi=0$, while in Refs.~\cite{pokrovskytalapov,Lazarides2009} (amongst numerous others), there is no such restriction on $\phi$. This is crucial to the position of the boundary of the commensurate-incommensurate transition, and is due to the fact that we are working at fixed particle number~\cite{giamarchi}.

\begin{figure}[t]
\centering \includegraphics[width=8cm]{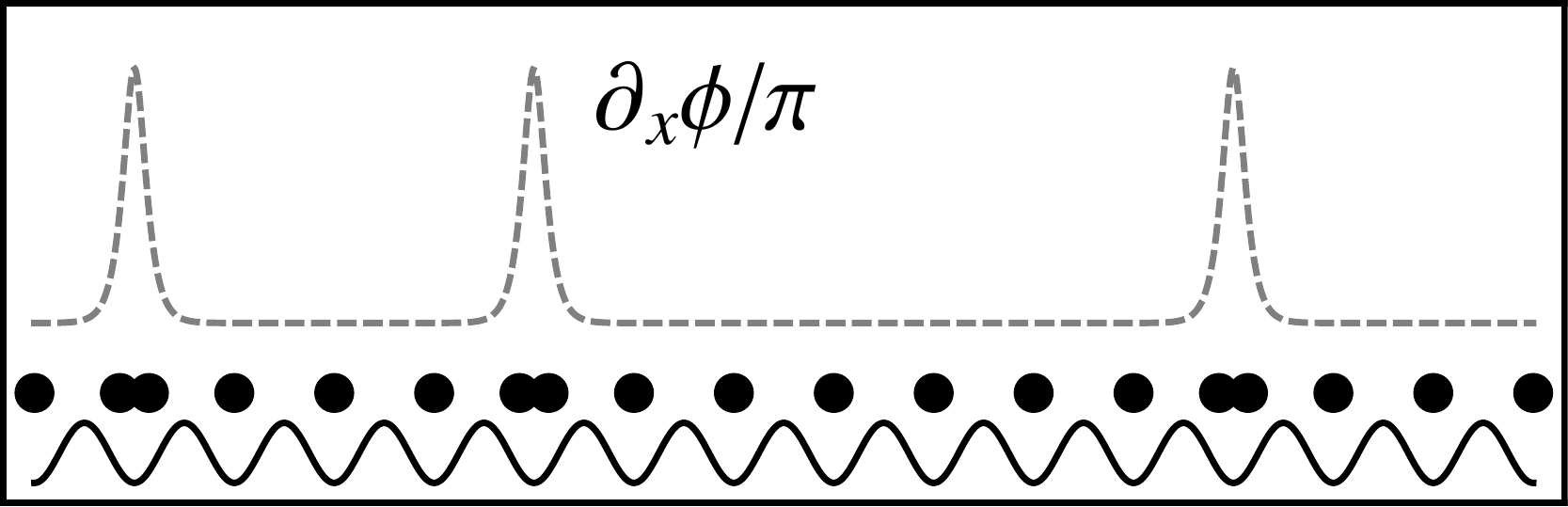} 
\caption{
	\label{figballsinlatticefinQ}
	$Q>0$; the dashed line is $\partial_x\phi/\pi$, the solid line is the periodic potential, and the corresponding particle positions are indicated by dots. Notice how kinks (indicated by localized deviations from a straight line for $\partial_x\phi/\pi$) correspond to particles in excess of the commensurate particle density.
}
\label{figballshighQ} 
\end{figure}

Since kinks correspond to excess particles above the commensurate density (see Fig.~\ref{figballshighQ}), fixing the particle density {\it must} fix the number density of kinks uniquely. But from its definition, $Q$ is directly proportional to this excess particle density, so that {\it the kink density must be proportional to $Q$ itself}. This immediately implies that $Q_c=0$, at least at zero temperature~\footnote{In Ref.~\cite{Buchler}, a finite value for $Q_c$ is obtained which, however, should be interpreted as a finite chemical potential, not density.}. Mathematically, this is a consequence of the boundary condition at the edges of the system, which implies that the commensurate phase cannot exist unless $Q=0$. For any $Q>0$, a finite density of bosonic kinks appears.

%\subsection{Inverse quadratic interactions}
The full analytical expression for the bare interaction between two kinks in a sine-Gordon model is given in Ref.~\cite{perringskyrme}; its limiting forms are
\begin{equation}
	V_{\mathrm{bare}}(r)\rightarrow
		\begin{cases}
			\frac{8}{\pi\xi}\exp(-r/\xi), &r\gg\xi\\
			\pi/(2r), &r\ll \xi.
		\end{cases}
	\label{eqpotnlimits}
\end{equation}
The effective width of the kinks is $\xi=1/\sqrt{2\pi K n_0 V}$, so that this potential amounts to an impenetrable core plus a finite-range repulsion. 

The effective kink (and antikink) mass $\sigma$ is strongly renormalized by quantum fluctuations~\cite{gogolin}. It may be obtained using a flow-equation RG scheme as described in Ref.~\onlinecite{kehrein}; the salient points are that it is proportional to $u$ at $K=1$ (that is, it is not renormalized) and vanishes as $K\rightarrow 2^-$. This vanishing of $\sigma$ is also responsible for the SF-MI transition observed for weak lattices in Ref~\cite{Chris} at commensurability; $K\rightarrow 2$ corresponds to a critical $\gamma_c\approx 3.5$. We obtain $\sigma$ for intermediate values $1<K<2$ by direct numerical integration of the flow equations given in Ref.~\cite{kehrein}.

To study the system of interacting kinks, we employ the statistical density matrix in imaginary time and position representation~\cite{ceperleyreview,feynmanhibbs}. This is given by
\begin{equation}
	\rho (R, R; \beta )=\int \prod _j\mathbb{D}x_j\exp
		(-S[\{x_n\}] / \hbar)
	\label{eqdensitymatrixwithstrings}
\end{equation}
where $\{x_n(\tau)\}$ denotes the set of positions of the particles at time $\tau$, $R=\{r_1,r_2,\ldots,r_N\}$ denotes the set of positions of the particles at $\tau=0$ and $\tau=\beta$ (see below) while $\mathbb{D}x_j$ denotes functional integration over $x_j$; finally,
\begin{equation}
	S[\{x_n\}]= \frac{1}{2} \int%_0^\beta 
	d\tau\,\left[
	\sigma\sum _n\left(\partial _{\tau}x_n\right)^2+\sum _{n,m}V\left(x_n-x_m\right)
		\right].
	\label{eqS}
\end{equation}
In Eq.~\eqref{eqS}, the integral runs from zero to $\beta$ and there is an ultraviolet cutoff $\Lambda_\tau=2\pi/\Delta \tau$ with $\Delta \tau$ a discretisation step size~\cite{ceperleyreview}. In this picture, the worldlines of the particles $x_i(\tau)$ correspond to classical strings without overhangs, the ends of which are fixed to $x_i(0)=x_i(\tau)=r_i$. Note that Eq.~\eqref{eqdensitymatrixwithstrings} and Eq.~\eqref{eqS} describe the $\rho$ appropriate for distinguishable particles; for bosonic particles, one symmetrises in the end, so that $\rho_B(R,R';\beta)=\sum_P \rho(R,P R';\beta)$, with $P$ labelling the permutation.

We begin by estimating the temperature dependence of the critical incommensurability $Q_d$ above which exchange effects become important. The worldlines of the particles are of length $\beta$ in the time-like direction, and the ``width'' of the path in the space-like direction will be $w\propto\sqrt{\hbar^2 \beta/\sigma}$. If the average inter-kink distance, proportional to $Q^{-1}$, is larger than this, quantum effects are not important; the condition for the statistics to be important is therefore $Q\sqrt{\hbar^2\beta/\sigma}>1$, up to a numerical factor. This defines a critical $Q_d\propto\sqrt{k_B T\sigma/\hbar^2}$. Below this $Q_d$, the kinks behave like noninteracting bosons; above it, they begin to interact, and we expect the effects described below to be evident.  Furthermore, since $\sigma$ vanishes on the lines $V=0$ and $K=2$, $Q_d$ also vanishes there.% see Fig~\ref{figfiniteT}. 

Next, we concentrate on the $T=0$ or $\beta\rightarrow\infty$ limit, corresponding to infinitely long strings; in this limit, the degeneracy condition is always satisfied. We shall employ a renormalization group (RG) technique applied directly to the density matrix of Eq.~\eqref{eqdensitymatrixwithstrings}. Details of this will be presented elsewhere~\cite{rgdetails}; here, we shall only outline our conclusions.

Splitting the fields $x_i$ into slow and fast parts as usual~\cite{skma}, it is possible to extend the Wegner-Houghton approach~\cite{wegner,lipowskyfisher} to the many-body case, obtaining the flow equation for the potential
\begin{equation}
	\partial_\epsilon V=V+\frac{1}{2}x\partial_x V
				+
				\hbar\Lambda_\tau
				\log\left(1+\frac{2}{\sigma\Lambda_\tau}
					\partial_{x}^2V\right).
	\label{eqflow}
\end{equation}
Notice that the coarse-graining is done in the $\tau$ direction; thus, information on lengthscales comparable to the kink density is still present in the fixed point-potentials.

The physics of the system is determined by the fixed point potentials of Eq.~\eqref{eqflow}. For bare (initial) potentials that diverge at the origin~\footnote{Note that the potential between a kink and an antikink does {\it not} diverge at the origin~\cite{perringskyrme}; thus, our analysis does not apply to the SF obtained at $K>2$.}, these may be determined numerically; for $x\gg\hbar/\Lambda_\tau \sigma$, their behaviour is
\begin{equation}
	V_{fp}=\frac{\hbar^2}{2\sigma}\frac{\lambda(\lambda-1)}{x^2},
	\label{eqVfp} 
\end{equation}
where we have written the coefficient of $x^{-2}$ as $\hbar^2\lambda(\lambda-1)/2\sigma$  in order to make contact with the conventional notation in the literature (see below).
We are thus dealing with a system of bosons interacting via an inverse square power law; this is the celebrated Calogero-Sutherland model~\cite{beautifulmodels}, the ground-state wavefunction and low-energy spectrum of which are known. We concentrate here on its ground-state properties, which have been studied using numerical techniques~\cite{astrakharchik}. The authors of Ref.~\cite{astrakharchik} find (quasi-)long-range off-diagonal order for $0<\lambda<2$, while they find (quasi-)long-range diagonal order for $\lambda>1$. The system is thus in a condensed, SF state for $0<\lambda<1$, in a SS state, characterised by the simultaneous presence of diagonal and off-diagonal long-range order for $1<\lambda<2$, and in a crystalline, S state characterised by strong diagonal correlations for $2<\lambda$. Therefore, the phase in which the system is for incommensurate densities ($Q\neq 0$) depends on the range in which the $\lambda$ corresponding to the potential in Eq.~\eqref{eqpotnlimits} lies.

To map out the phase diagram, we note that local analysis of the fixed point ordinary differential equation, Eq.~\eqref{eqflow} with the left hand side set to 0, indicates that the fixed point potentials, $V_{fp}$, have the property that $\partial V_{fp}/\partial\sigma>0$ (for all $x$). An increase in $\sigma$ therefore results in an increase in $\lambda$ of Eq.~\eqref{eqVfp}. In addition, at $K=1$ and $V=0$ (hard-core free bosons), $\lambda=1$~\cite{Cazalilla}. Based on these two pieces of information, and the behaviour of the mass described earlier, we propose the phase diagram in Fig.~\ref{figphasediagT0} for $T=0$. Starting from the point $K=1$, $V=0$, an increase of $V$ causes a rapid increase of $\sigma$, which corresponds to an increase in $\lambda$ so that $\lambda>1$ which corresponds to SS. As $V$ is further increased, $\lambda$ reaches the value $\lambda=2$ at $V=V_{c,m}$, at which point phase coherence is lost, the structure factor displays a sharp peak~\cite{astrakharchik}, and the system is in the S state. \ignore{This S state is unusual in that the filling factor is not an integer; it corresponds to the kinks forming a lattice, and is not directly related to the underlying lattice.} On the other hand, starting from any point on the $K=1$ line and increasing $K$ corresponds to decreasing $\sigma$, thus decreasing $\lambda$ from its value at $K=1$. As a result, the line $V_{c,m}$ curves upwards as $K$ increases. Starting from $V_{c,0}(K=1)=0$ and increasing $K$, $\lambda$ must decrease below 1 so that the diagonal order is suppressed; thus, the line $V_{c,0}$ separating the SF from the SS region also curves upwards. As $K\rightarrow 2$, or $\gamma\rightarrow 3.5$, the effective mass vanishes for any $V$; this results in a rapid decrease of $\lambda$, so that both lines curve upwards sharply.

It is important to note that the presence of the SS phase represents an order out of disorder effect: quantum fluctuations, which at first sight one would expect disorder the system, result in a strengthening of the repulsion which in turn causes the system to order.

Let us briefly discuss the differences between the phases just described in terms of experimentally accessible quantities. The main distinguishing features of these phases are the diagonal and off-diagonal correlations. Off-diagonal long-range order may be observed using time-of-flight measurements, which therefore allow us to distinguish the SF and SS phase-coherent phases from the S phase; in the latter, phase correlations drop quickly and the time-of-flight image is smeared. On the other hand, techniques for measuring density variations would distinguish between the SS and S phases on one hand and the SF phase on the other; single-site addressability is possible~\cite{gericke}, which  may alllow to detect density oscillations. 

%\section{Conclusions}

In summary, we have shown that the incommensurability-induced MI-SF transition occurs for arbitrarily small incommensurability. We have then studied the system of bosonic quasiparticles which appears as soon as commensurability is lost; calculating the effective interactions between them, as well as their effective mass, and using an RG transformation, we have argued that quantum fluctuations drive the interactions to acquire an inverse square form (a Calogero-Sutherland model). A phase diagram for the current system of strongly interacting bosons in a weak 1D optical lattice is then proposed, which features SF, SS, and S phases. The most striking feature of our calculations is the appearance of a SS phase, in spite of the local character of the interactions in the original model. The periodicity of both the SS and the S phases is unrelated to that of the underlying lattice, thus providing us with more exotic states of matter.

Although the experimental setup of Ref.\ [\onlinecite{Chris}] allows, in principle, to tune the density and to investigate also the commensurate-incommensurate quantum phase transition, up to now only the commensurate regime has been studied. Our studies indicate that the elusive SS phase is within reach by modifying a single parameter in the experimental setup of Ref.\ [\onlinecite{Chris}]. We hope that our work will trigger further experiments into this fascinating  and largely unexplored regime.

We acknowledge financial support from the Netherlands Organization for Scientific Research (NWO), and thank J.-S. Caux, G. Japaridze and L. K. Lim for stimulating discussions and for a critical reading of the manuscript.

\end{document}